\documentclass[useAMS,usenatbib,usegraphicx]{mn2e}

\title[Internal motions in OB-associations] {Internal motions in OB-associations with {\it Gaia} DR2}
\author[A.M. Melnik and A.K. Dambis]{A.M.Melnik\thanks{E-mail: anna@sai.msu.ru}   and A.K.
Dambis\\ Sternberg Astronomical Institute, Lomonosov Moscow State
University, Universitetskii pr. 13, Moscow, 119991 Russia}

\begin{document}

\date{Accepted 2020 February 11. Received 2020 January 17; in original form 2019 June 6}


\maketitle

\label{firstpage}

\begin{abstract}

We study the motions inside  28 OB-associations with the use of
{\it Gaia} DR2 proper motions. The average velocity dispersion
calculated for 28 OB-associations including more than 20 stars
with {\it Gaia} DR2 proper motion  is  $\sigma_v =4.5$ km
s$^{-1}$. The median virial and stellar masses of OB-associations
are $M_{vir}=8.9 \times 10^5$  and $M_{st}=8.1 \times 10^3$
M$_\odot$, respectively. The median star-formation efficiency in
parent giant molecular clouds appears   to be $\epsilon=1.2$ per
cent. {\it Gaia} DR2 proper motions confirm the expansion in the
Per OB1, Car OB1 and Sgr OB1 associations found earlier with {\it
Gaia} DR1 data. We also detect the expansion in Gem OB1, Ori OB1
and Sco OB1   associations which became possible for the first
time now when analyzed  with {\it Gaia} DR2 proper motions. The
analysis of the distribution of OB-stars in the Per OB1
association shows the presence of a shell-like structure with the
radius of 40 pc. Probably, the expansion of the Per OB1
association started with the velocity greater than the present-day
expansion velocity equal to $5.0\pm1.7$ km s$^{-1}$.

\end{abstract}


\section{Introduction}

The second intermediate {\it Gaia} data release ({\it Gaia} DR2),
which is based on the data collected during the first 1.8 yr of
{\it Gaia} mission, provides proper motions  for more than 1.3
billion stars with a characteristic  accuracy of $\sim0.1$ mas
yr$^{-1}$ \citep{brown2018,lindegren2018,katz2018}. The main
achievement of {\it Gaia} DR2  is the large number of stars with
high-precision astrometric data.

The first {\it Gaia} data release ({\it Gaia} DR1)  contains {\it
Tycho-{\it Gaia}} Astrometric Solution
\citep[TGAS,][]{michalik2015,lindegren2016} for $\sim2$ million
stars based on positions spanning a 24 yr time interval between
Hipparcos \citep{esa1997} and {\it {\it Gaia}} \citep{gaia2016a}
measurements.

\citet{melnik2017}   used {\it Gaia} DR1 (TGAS) data to identify
500 stars in OB-associations. These stars have their first-epoch
positions determined by the Hipparcos catalog \citep{esa1997},
which provides the accuracy of TGAS proper motions of $\sim0.06$
mas yr$^{-1}$. The average one-dimensional velocity dispersion
inside 18 OB-associations with more than 10 TGAS stars appeared to
be $\sigma_v=3.9$ km s$^{-1}$. Precise {\it Gaia} DR1 (TGAS)
proper motions allowed us to find the expansion in the Per OB1,
Car OB1 and Sgr OB1 associations determined at significance level
$P>2.5\sigma$ \citep{melnik2017}.

As for the motion of OB-associations as whole entities, the
results obtained with {\it Gaia} DR1 and {\it Gaia} DR2 data are
in good agreement. The median velocities of OB-associations
derived from {\it Gaia} DR1 and {\it Gaia} DR2 proper motions
differ on average by 2 km s$^{-1}$. The parameters of the rotation
curve calculated with proper motions from {\it Gaia} DR1 and {\it
Gaia} DR2  are consistent within the errors
\citep{melnik2017,melnik2019}.

OB-associations are sparse groups of O and B stars
\citep{ambartsumian1949}. They are supposed to be born in giant
molecular clouds \citep{elmegreen1983, zinnecker2007}. There is
extensive evidence that  giant molecular clouds are in a state
close to virial equilibrium
\citep{larson1981,krumholz2006,kauffmann2013, chen2019}. The
expected sizes and masses of giant molecular clouds are 10--80 pc
and 10$^5$ -- 2 10$^6$ M$_\odot$, respectively
\citep{sanders1985}. The estimates of the average efficiency of
star formation, $\epsilon$, in giant molecular clouds defined as
the ratio of the total mass of stars born inside a cloud to the
initial  gas mass lies in the range of 0.1--10 percent
\citep{myers1986, evans2009, garcia2014}. The median
star-formation efficiency determined for 18 OB-associations with
more than 10 TGAS stars appears to be 2.1 per cent
\citep{melnik2017}.

The radiation of massive  stars creates HII regions, which can
destroy molecular clouds \citep{mckee1989, franco1994,
colin2013,kim2016}. If a gas cloud loses more than 50 per cent of
its mass in a time less than one crossing time, then the system
becomes unbound \citep{hills1980}. But if the mass is ejected
slowly then the system can form an expanding OB-association with a
bound cluster in its centre
\citep{kroupa2001,boily2003a,boily2003b,baumgardt2007}.

The catalog by  \citet{blahahumphreys1989} comprises 91
OB-association located within $\sim3$ kpc of the Sun.
\citet{melnik1995} found that many  of OB-associations identified
by \citet{blahahumphreys1989}   include several centres of
concentration.

The study by \citet{blaauw1964} stimulated the interest in the
search of expanding OB-associations.  Several  methods  have been
proposed for determining the parameters of expansion in stellar
groups \citep{brown1997,madsen2002}.   Many studies of expansion
of OB-associations based on  {\it Gaia} data have been undertaken
in the last few years. \citet{kounkel2018} studied the kinematics
of the Ori OB1 association and found  significant expansion in
only one subgroup inside it. \citet{cantat2019} studied the Vela
OB2 association and detected  expansion in all seven detected
subgroups. \citet{wright2018} investigated the Sco OB2 association
and found no evidence of expansion inside three selected
subgroups. \citet{ward2018} identified 18 associations on the
basis of their  original method and reported that none of them
shows evidence of expansion.

Here we do not use {\it Gaia} DR2 parallaxes for stars of
OB-associations because they  seem to need some correction of
their zero point, moreover, different studies give different
values for this correction \citep[][and other
papers]{lindegren2018, arenou2018, stassun2018, riess2018,
yalyalieva2018}. Exception is the Ori OB1 association for which
trigonometric and photometric methods give the same distance of
0.4 kpc.  Note that associations located at the distances of $\sim
2$ kpc  do not show such an agreement. We also do not consider
{\it Gaia} DR2 line-of-sight velocities because they are measured
for only 7 per sent of stars of OB-associations.

In this paper we study the internal motions of young stars inside
OB-associations as derived from  {\it Gaia} DR2 proper motions.
Section 2 describes the catalog of stars in OB-associations with
{\it Gaia} DR2 astrometric and photometric data. Section 3
analyzes  velocity dispersion inside OB-associations, determines
the virial and stellar masses of OB-associations, estimates the
star-formation efficiency in giant molecular clouds,  studies the
expansion of OB-associations and   the role of shell-like
structures. Section 4  discusses the results and formulates the
main conclusions.

\section{Data}

The catalog by  \citet{blahahumphreys1989}  includes 2209
high-luminosity  stars of OB-associations, 2007 of them  have been
identified with {\it Gaia} DR2 catalog and 1990  (90 per cent)
have  {\it Gaia} DR2 proper motions.

We identified members of OB-associations from the catalog by
\citet{blahahumphreys1989} with {\it Gaia} DR2 stars  based on the
proximity of stellar coordinates and  visual magnitudes. The
coordinates in the catalog by \citet{blahahumphreys1989} are given
with small accuracy and therefore  we also invoke data from
Strasbourg astronomical Data Center.

We  cross-matched the list of \citet{blahahumphreys1989} with {\it
Gaia} DR2 catalog using  a matching radius of 3 arcsec and the
magnitude tolerance of $|G_{BH}-G|<3^m$ which gave us matches for
a total of 2007 stars. Out of them the 3-arcsec matches were
unique for 1873 stars, 131 stars had two matches, and 3 stars had
three matches. For stars with unique matches the matching distance
did not exceed 0.5 arcsec (the median value is 0.051 arcsec), for
stars with 2 or 3 matches the distance for the closest match did
not exceed 0.19 arcsec (the median value is 0.048 arcsec), the
median distances to the second-and third matches is 2.01 arcsec,
respectively. We therefore adopted all matches where they were
unique and adopted the closest match in the remaining cases.

We also compare {\it Gaia} $G$-band magnitudes with the $G$-band
magnitudes ($G_{BH}$) predicted for stars from
\citet{blahahumphreys1989} list based on the $B$- and $V$-band
data provided in it.  The predicted $G$-band values were derived
on the basis of an empirical relation involving ($B-V$) color
indices and $V$-band magnitudes \citep{jordi2010}:
$$
G_{BH} = V-0.0424-0.0851\,(B-V)-0.3348\,(B-V)^2
$$
\begin{equation}
\hspace{3.1 cm}+0.0205\,(B-V)^3.
 \label{gbh}
\end{equation}

Table~1 (available in the online version of the paper) lists the
kinematic and  photometric data for stars in OB-associations
obtained with {\it Gaia} DR2. It presents the name of a star, the
name of the OB-association to which it is assigned, spectral type
of the star, its  luminosity class and  visual magnitude, $m_v$,
which all are taken from the catalog   by
\citet{blahahumphreys1989}. We also show the heliocentric distance
to the OB-association by \citet{blahahumphreys1989}, $r_{BH}$,
reduced to the short distance scale, $r=0.8\, r_{BH}$
\citep{sitnik1996,dambis2001, melnikdambis2009}. Table~1 also
represents  {\it Gaia} DR2 data: equatorial coordinates, $\alpha$
and $\delta$, of the star, its Galactic coordinates, $l$ and $b$,
the magnitude in the $G$-band, the parallax, $\pi$, proper motions
along  $l$- and $b$-directions, $\mu_l$ and $\mu_b$, and their
errors, $\varepsilon_{\pi}$, $\varepsilon_{\mu_l}$ and
$\varepsilon_{\mu_b}$.  It also lists the number of visibility
periods, $n_{vis}$, of the star, i.~e. the number of groups of
observations separated from each other  by at least 4 days
\citep{arenou2018}, and the re-normalised unit weight errors
(RUWE), which measure how well the {\it Gaia} observations agree
with the five-parameter single-star model and whose large value
(RUWE$>1.4$) could indicate an astrometric binary
\citep{lindegren2018a,lindegren2019}.   For completeness we also
added to Table 1 the stellar line-of-sight velocities, $V_r$, and
their errors, $\varepsilon_{vr}$, taken from the catalog by
\citet{barbierbrossat2000}, which are available for 52 per cent of
stars of OB-associations.  Here we will consider the refined
sample  of stars in OB-associations including 1771 stars with
$n_{vis}>8$ and RUWE$<1.4$. Of 219 excluded stars, 174 stars have
RUWE$\ge1.4$ and 45 objects have $n_{vis}\le 8$.

\addtocounter{table}{1}

\section{Results}

\subsection{Velocity dispersion inside  OB-associations}

Let us consider the motions of stars in OB-associations in the sky
plane. The velocity components of a star along $l$- and
$b$-directions, $v_l$ and $v_b$, are calculated as follows:

\begin{equation}
v_l = 4.74\, \mu_l \, r,
\end{equation}
\begin{equation}
v_b = 4.74 \,\mu_b \, r,
\end{equation}

\noindent where $r$ is the heliocentric distance of the
association but $\mu_l$ and $\mu_b$ are proper motions in mas
yr$^{-1}$.  The factor $4.74\times r$~(kpc) transformers  units of
mas yr$^{-1}$ into km s$^{-1}$.

We determine the standard deviations, $\sigma_{vl}$ and
$\sigma_{vb}$, of velocity components, $v_l$ and $v_b$, in an
association as half of the velocity interval, $\Delta v_l$ and
$\Delta v_b$, including  central 68 per cent of member stars with
known {\it Gaia} DR2 proper motions. These robust estimates of the
velocity dispersion allow us to minimize the contribution of
outliers.

The average uncertainty in determination of proper motions of
stars of OB-associations in {\it Gaia} DR2 is 0.086 mas yr$^{-1}$,
which corresponds to the uncertainty in the sky-plane velocity of
stars at the distance of $r=1$ kpc equal to 0.4 km s$^{-1}$.

\begin{figure}
\resizebox{\hsize}{!}{\includegraphics{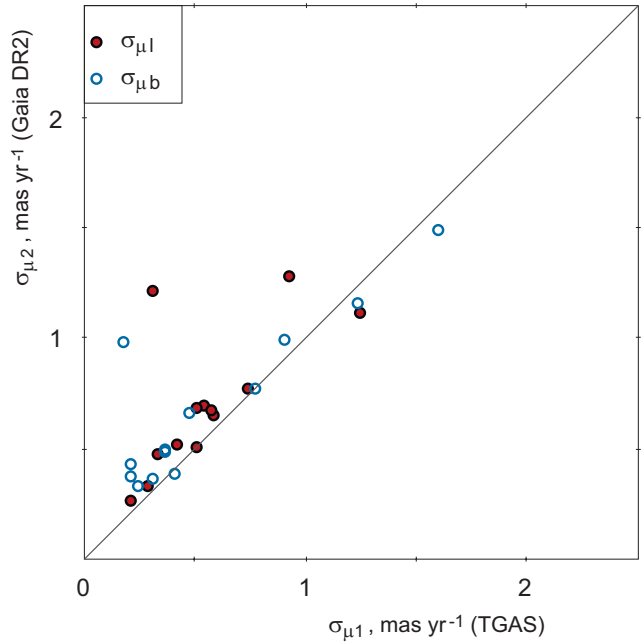}}
\caption{Comparison of the dispersions of the proper motions of
stars inside OB-associations derived with {\it Gaia} DR2  and with
{\it Gaia} DR1 (TGAS) data. The solid line shows the bisectrix. We
can see that the dispersions obtained with {\it Gaia} DR2 proper
motions  are systematically larger than those calculated with {\it
Gaia} DR1. The  Cyg OB7 association with the dispersions
$\sigma_{\mu 1}=3.76$ and $\sigma_{\mu 2} =4.51$ mas yr$^{-1}$ is
located beyond the limits of the plot.} \label{com}
\end{figure}

Figure~\ref{com}  compares  the dispersions of stellar proper
motions inside OB-associations derived with {\it Gaia} DR1 (TGAS)
and {\it Gaia} DR2 data, $\sigma_{\mu 1}$ and $\sigma_{\mu 2}$,
respectively.  We can see that the dispersions obtained with {\it
Gaia} DR2 proper motions are systematically larger than those
calculated with {\it Gaia} DR1 data. The linear relation between
them is $\sigma_{\mu 2}=1.09\pm0.08\;\sigma_{\mu 1}$. Here we
consider 13 OB-associations with $\sigma_{\mu 2}<2.5$ mas
yr$^{-1}$ and including more than 10 stars  common for the
Hipparcos, {\it Gaia} DR1 and {\it Gaia} DR2 catalogs.

Table~\ref{disper} lists the average  standard deviations,
$\overline{\sigma_{vl}}$ and $\overline{\sigma_{vb}}$, of
velocities, $v_l$ and $v_b$, inside OB-associations, calculated
for different samples of member stars with  {\it Gaia} proper
motions.

We corrected the velocity dispersions inside OB-associations for
the inflationary-effect of measurement errors by the following
way:

\begin{equation}
\sigma_{vl}^2=\sigma_{l,\,obs}^2-(4.74\; r\, \varepsilon_{\mu
l})^2,
\end{equation}
\begin{equation}
\sigma_{vb}^2=\sigma_{b,\,obs}^2-(4.74\; r\, \varepsilon_{\mu
b})^2,
\end{equation}

\noindent where $\varepsilon_{\mu l}$ and $\varepsilon_{\mu b}$
are the average  errors in determination of proper motions inside
an association in the $l$- and $b$-directions, respectively.
Table~\ref{disper} also presents the average  velocity dispersions
derived for both $l$- and $b$-directions:

\begin{equation}
\overline{\sigma_v}=(\overline{\sigma_{vl}}+\overline{\sigma_{vb}})/2;
\end{equation}

\noindent and  the root-mean-square errors in determination of the
average velocity dispersions. Table~\ref{disper} also lists the
minimal number of stars, $n_\mu$, with known {\it Gaia} proper
motions, which an association must include for it to be considered
in our study, and the number of associations in the sample, $n$.
The comments indicate the criteria for selection of stars in
associations. The first four rows are related to {\it Gaia} DR2
proper motions while the fifth row presents quantities derived
from {\it Gaia} DR1 data. The first and second rows represent
samples including associations with more than $n_\mu>20$ and
$n_\mu>10$ {\it Gaia} DR2 stars, respectively. The third, fourth
and fifth rows lists quantities  derived only for stars common for
Hipparcos and {\it Gaia} catalogs.  The selection of Hipparcos
stars in {\it Gaia} DR1 (TGAS) data allows us to use the most
precise proper motions (fifth row).

Table~\ref{disper}  suggests that all estimates of the velocity
dispersions inside OB-associations derived from {\it Gaia} DR2
data (the first four rows) are greater than those obtained with
{\it Gaia} DR1 (the bottom row). However, the statistical
significance of this predominance is not large (just $P\sim
1\,\sigma$). For example, the average  velocity dispersion
$\overline{\sigma_v}$ calculated for 28 associations with  more
than  $n_\mu>20$ {\it Gaia} DR2 proper motions (first row) amounts
to 4.46$\pm0.30$ km s$^{-1}$ but $\overline{\sigma_v}$ computed
for {\it Gaia} DR1 data is  only 3.54$\pm0.49$ km s$^{-1}$ (fifth
row). Note that the velocity dispersions shown in the fourth and
fifth rows are derived on the basis of the same sample of stars,
and their comparison gives the most convincing argument that the
velocity dispersion, $\overline{\sigma_{v}}$, obtained with {\it
Gaia} DR2 data is greater than $\overline{\sigma_{v}}$ calculated
with {\it Gaia} DR1 data at a significance level of $P\sim
1\sigma$.

We excluded Cep OB1 ($r=2.8$ kpc), NGC 2439 ($r=3.5$ kpc) and R
103 ($r=3.2$ kpc) associations from consideration because of their
distant location from the Sun ($r>2.8$ kpc), which results in
increased probability of wrong membership. Moreover, the Cep OB1
and NGC 2439 associations are very elongated in the l-direction,
the ratio of their sizes in the l- and b-direction exceeds
$d_l/d_b>2$ which suggests that they can include a chain of
OB-associations. The R 103 association has too large velocity
dispersion, $\sigma_{vl}=21$ km s$^{-1}$, and seems to be strongly
contaminated by field stars.

Generally, the observed velocity dispersion inside OB-associations
has several sources:  turbulent motions inside giant molecular
clouds, in which young stars were born;  motions inside binary
systems; the uncertainty in the determination of  proper motions.
However, the contribution of the latter source is insignificant.

Here we suppose that the contribution of binary systems into the
velocity dispersion inside OB-associations is small, no greater
than $\sigma_{bn} \sim 1$ km s$^{-1}$. Some estimates of the
binary effect inside OB-associations will be presented in
separated paper \citep{melnik2020}.

\begin{table*}
\caption{Average velocity dispersion inside OB-associations}
 \begin{tabular}{llcccccl}
 \\[-7pt] \hline\\[-7pt]
Row & Catalog & Condition & $\overline{\sigma_{vl}}$ & $\overline{\sigma_{vb}}$ & $\overline{\sigma_{v}}$ & $n$ & Comments\\
[2pt]
&& &    (km s$^{-1}$)   &   (km s$^{-1}$) & (km s$^{-1}$) &   \\
\\[-7pt] \hline\\[-7pt]
1 &{\it Gaia} DR2 & $n_\mu>20$ & 4.96$\pm 0.49$ & 3.96$\pm 0.36$ & 4.46$\pm 0.30$ & 28 & Stars identified with {\it Gaia} DR2 \\
2 &{\it Gaia} DR2 & $n_\mu>10$ & 4.65$\pm 0.39$ & 4.69$\pm 0.33$ & 4.17$\pm 0.26$ & 40 & Stars identified with {\it Gaia} DR2 \\
3 &{\it Gaia} DR2 & $n_\mu>10$ & 4.69$\pm 0.49$ & 3.96$\pm 0.47$ & 4.33$\pm 0.34$ & 23 & Common stars for {\it Gaia} DR2 and Hipparcos \\
4 &{\it Gaia} DR2 & $n_\mu>10$ & 5.30$\pm 0.88$ & 4.42$\pm 0.81$ & 4.86$\pm 0.60$ & 14 & Common stars for {\it Gaia} DR1,  DR2 and Hipparcos\\
5 &{\it Gaia} DR1 & $n_\mu>10$ & 3.77$\pm 0.64$ & 3.30$\pm 0.75$ & 3.54$\pm 0.49$ & 16 & Common stars for {\it Gaia} DR1 and Hipparcos   \\
\\[-7pt] \hline\\[-7pt]
\end{tabular}
\label{disper}
\end{table*}

\subsection{Efficiency of star formation}

Giant molecular clouds, from which OB-associations  form, seem to
be close to their virial equilibrium.  The velocity dispersion of
turbulent motions, $\sigma_t$, inside OB-association and the
radius of association, $a$,  can be used to estimate the virial
masses, $M_{vir}$, of OB-associations, which are equal to the
masses of their parent molecular clouds:

\begin{equation}
 M_{vir}=\frac{5a\sigma_t^2}{G},
\label{mvir}
\end{equation}

\noindent  where $a$ is the specific radius of an association.
Here we  assume $a$ to be the radius containing central 68 per
cent of association member stars. The velocity dispersion of
turbulent motions, $\sigma_t$, is supposed to be close to the
observed velocity dispersion corrected for measurement errors,
$\sigma_v$:

\begin{equation}
 \sigma_t \approx \sigma_v.
\label{sig_t}
\end{equation}

Another remark is connected with the expansion of the  Car OB1 and
Per OB1 associations,   which had a considerably smaller size some
time ago -- this seems to be their initial size -- so their
present-day radius $a$ must be corrected by a factor:

\begin{equation}
 a_c=a\,\xi,
\label{ac}
\end{equation}

\noindent where $\xi$ is the ratio of the minimum to the
present-day radius of an association (see section 3.4,
Eq.~\ref{size}).

We can also calculate the stellar masses of OB-associations,
$M_{st}$, through the  number of stars with masses greater than 20
M$_\odot$, $N_{20}$.  Note that  \citet{blahahumphreys1989}
considered the catalog of stars in OB-associations to be fairly
complete for stars brighter than $M_{bol}<-7.5^m$, which
corresponds to stars with masses greater than $\sim20$ M$_\odot$
\citep{bressan2012}. The power-law mass function by
\citet{kroupa2002} calibrated via $N_{20}$ can be used to estimate
the full mass of stars in an association \citep[for more details
see][]{melnik2017}.

Table~\ref{mass} lists the virial and stellar masses, $M_{vir}$
and $M_{st}$, of OB-associations containing  more than 20  stars
with {\it Gaia} DR2 proper motions, $n_\mu>20$. It also lists the
general parameters of an OB-association: the average Galactic
longitude and latitude, $l$ and $b$; the average heliocentric
distance, $r$;  the median line-of-sight velocity, $V_r$, and the
number of stars  it is derived from given in parentheses;  the
specific radius of OB-association, $a$; the number of stars with
masses $M>20$ M$_\odot$, $N_{20}$; the velocity dispersions in
$l$- and $b$-directions, $\sigma_{vl}$ and $\sigma_{vb}$, and the
number of stars  with known {\it Gaia} DR2 proper motions,
$n_\mu$. Also listed in Table~\ref{mass} is the average efficiency
of star formation, $\epsilon$, inside the parent giant molecular
cloud:

\begin{equation}
\epsilon=M_{st}/M_{vir}.
 \label{epsilon}
\end{equation}

Table~\ref{mass} indicates that the star-formation efficiency in
OB-associations ranges from  0.1 to 15 per cent. Its median value
calculated for 28 OB-associations  is $\epsilon=1.2$ per cent. The
median  virial and stellar masses appear to be $8.9 \times 10^5$
and $8.1 \times 10^3$ M$_\odot$, respectively.

\begin{table*}
\caption{Virial and stellar masses of OB-associations, $M_{vir}$
and $M_{st}$, and star formation efficiency $\epsilon$}
 \begin{tabular}{lrrccccrrrrrr}
 \\[-7pt] \hline\\[-7pt]
Name & $l\quad$ & $b \;\;\;$ & $r$  & $V_r$ & $\sigma_{vl} $ &
$\sigma_{vb}$ &  $a$  & $n_\mu$ &
$M_{vir}\quad$ & $M_{st}\quad$ & $N_{20}$ & $\epsilon$ 100\% \\
[2pt]
& (deg) & (deg) & (kpc)  &   (km s$^{-1})$   &(km s$^{-1})$   &  (km s$^{-1})$ & (kpc)  & &  $M_\odot\quad$ &   $M_\odot\quad$ &  &   \\
  \\[-7pt] \hline\\[-7pt]
SGR OB5   &  0.04 & -1.16 &  2.42 & -15.0 ( 2) &13.4 & 7.0 &0.068 &  27 &    83.2 $\times 10^5$ &    7.5 $\times 10^3$ &  18 &  0.1$\;\;$  \\
SGR OB1   &  7.54 & -0.77 &  1.26 & -10.0 (37) & 2.5 & 5.0 &0.037 &  47 &     6.0 $\times 10^5$ &    9.2 $\times 10^3$ &  22 &  1.5$\;\;$  \\
SER OB1   & 16.71 &  0.07 &  1.53 &  -5.0 (17) & 4.0 & 4.0 &0.051 &  33 &     9.7 $\times 10^5$ &    8.3 $\times 10^3$ &  20 &  0.9$\;\;$  \\
CYG OB3   & 72.76 &  2.04 &  1.83 & -10.0 (29) & 6.8 & 4.1 &0.024 &  32 &     8.5 $\times 10^5$ &   10.8 $\times 10^3$ &  26 &  1.3$\;\;$  \\
CYG OB1   & 75.84 &  1.12 &  1.46 & -13.5 (34) & 6.2 & 3.6 &0.032 &  62 &     9.0 $\times 10^5$ &   16.7 $\times 10^3$ &  40 &  1.8$\;\;$  \\
CYG OB9   & 77.81 &  1.80 &  0.96 & -19.5 (10) & 3.4 & 3.4 &0.017 &  22 &     2.3 $\times 10^5$ &    4.6 $\times 10^3$ &  11 &  2.0$\;\;$  \\
CYG OB8   & 77.92 &  3.36 &  1.83 & -21.0 ( 9) & 4.7 & 9.5 &0.040 &  20 &    23.5 $\times 10^5$ &    6.7 $\times 10^3$ &  16 &  0.3$\;\;$  \\
CYG OB7   & 88.98 &  0.03 &  0.63 &  -9.4 (21) &11.0 & 2.2 &0.051 &  22 &    25.7 $\times 10^5$ &    2.9 $\times 10^3$ &   7 &  0.1$\;\;$  \\
CEP OB2   &102.02 &  4.69 &  0.73 & -17.0 (36) & 5.3 & 4.7 &0.046 &  45 &    13.5 $\times 10^5$ &    7.9 $\times 10^3$ &  19 &  0.6$\;\;$  \\
CAS OB2   &111.99 & -0.00 &  2.10 & -50.1 ( 7) & 7.8 & 6.5 &0.056 &  30 &    33.4 $\times 10^5$ &    9.6 $\times 10^3$ &  23 &  0.3$\;\;$  \\
CAS OB5   &116.09 & -0.50 &  2.01 & -45.8 (16) & 3.6 & 3.7 &0.043 &  45 &     6.7 $\times 10^5$ &    9.6 $\times 10^3$ &  23 &  1.4$\;\;$  \\
CAS OB4   &120.05 & -0.30 &  2.30 & -37.0 ( 7) & 7.8 & 5.3 &0.070 &  24 &    34.9 $\times 10^5$ &    5.4 $\times 10^3$ &  13 &  0.2$\;\;$  \\
CAS OB7   &122.98 &  1.22 &  2.01 & -50.0 ( 4) & 3.7 & 2.4 &0.041 &  35 &     4.6 $\times 10^5$ &    6.7 $\times 10^3$ &  16 &  1.5$\;\;$  \\
CAS OB8   &129.16 & -1.06 &  2.30 & -34.6 (14) & 1.7 & 2.1 &0.037 &  41 &     1.6 $\times 10^5$ &    8.3 $\times 10^3$ &  20 &  5.3$\;\;$  \\
PER OB1   &134.70 & -3.14 &  1.83 & -43.2 (80) & 3.7 & 2.9 &0.064 & 150 &     5.7 $\times 10^5$ &   36.2 $\times 10^3$ &  87 &  6.4$^a$  \\
CAS OB6   &134.95 &  0.72 &  1.75 & -42.6 (12) & 4.6 & 5.9 &0.057 &  29 &    18.5 $\times 10^5$ &    6.7 $\times 10^3$ &  16 &  0.4$\;\;$  \\
CAM OB1   &141.08 &  0.89 &  0.80 & -11.0 (30) & 3.3 & 2.9 &0.079 &  41 &     8.8 $\times 10^5$ &    5.0 $\times 10^3$ &  12 &  0.6$\;\;$  \\
AUR OB1   &173.83 &  0.14 &  1.06 &  -1.9 (26) & 4.0 & 3.0 &0.072 &  31 &    10.2 $\times 10^5$ &    3.7 $\times 10^3$ &   9 &  0.4$\;\;$  \\
ORI OB1   &206.90 &-17.71 &  0.40 &  25.4 (62) & 2.9 & 1.7 &0.035 &  54 &     2.2 $\times 10^5$ &    3.3 $\times 10^3$ &   8 &  1.5$\;\;$  \\
GEM OB1   &188.96 &  2.22 &  1.21 &  16.0 (18) & 3.8 & 2.5 &0.042 &  35 &     4.8 $\times 10^5$ &    5.4 $\times 10^3$ &  13 &  1.1$\;\;$  \\
MON OB2   &207.35 & -1.60 &  1.21 &  23.0 (25) & 3.3 & 4.0 &0.037 &  23 &     5.6 $\times 10^5$ &    5.4 $\times 10^3$ &  13 &  1.0$\;\;$  \\
VELA OB1  &264.83 & -1.41 &  1.46 &  23.0 (18) & 4.1 & 3.5 &0.057 &  43 &     9.5 $\times 10^5$ &   10.4 $\times 10^3$ &  25 &  1.1$\;\;$  \\
CAR OB1   &286.45 & -0.46 &  2.01 &  -5.0 (39) & 7.6 & 3.2 &0.064 & 101 &    15.5 $\times 10^5$ &   20.8 $\times 10^3$ &  50 &  1.3$^a$  \\
CAR OB2   &290.39 &  0.12 &  1.83 &  -8.2 (22) & 3.8 & 2.6 &0.028 &  48 &     3.3 $\times 10^5$ &   10.4 $\times 10^3$ &  25 &  3.1$\;\;$  \\
CRU OB1   &294.87 & -1.06 &  2.01 &  -5.3 (33) & 4.1 & 2.5 &0.040 &  65 &     5.1 $\times 10^5$ &    9.6 $\times 10^3$ &  23 &  1.9$\;\;$  \\
CEN OB1   &304.14 &  1.44 &  1.92 & -19.0 (32) & 5.1 & 2.9 &0.068 &  85 &    12.5 $\times 10^5$ &   20.4 $\times 10^3$ &  49 &  1.6$\;\;$  \\
ARA OB1A  &337.68 & -0.92 &  1.10 & -36.3 ( 8) & 4.6 & 7.8 &0.046 &  42 &    20.6 $\times 10^5$ &    4.6 $\times 10^3$ &  11 &  0.2$\;\;$  \\
SCO OB1   &343.72 &  1.37 &  1.53 & -28.8 (28) & 2.1 & 2.3 &0.013 &  66 &     0.8 $\times 10^5$ &   11.7 $\times 10^3$ &  28 & 15.0$\;\;$  \\

\\[-7pt] \hline\\[-7pt]
\multicolumn{12}{l}{ {\it Note.}$^a$ Values of $M_{vir}$ and
$\epsilon$ for the  Per OB1 and Car OB1 associations are corrected
for the expansion effect (Section 3.3)} \\
\end{tabular} \label{mass}
\end{table*}

\subsection{Expansion of OB-associations}

We determine possible  expansion or compression of OB-associations
via  parameters $p_l$ and $p_b$:

\begin{equation}
 v_l = v_{l0}+ p_l \; r \sin(l-l_0),
 \label{pl}
\end{equation}
\begin{equation}
 v_b = v_{b0}+ p_b \; r \sin(b-b_0),
 \label{pb}
\end{equation}

\noindent where $v_{l0}$ and $v_{b0}$ are the average velocities
of the association; $l_0$ and $b_0$ are the coordinates of  the
centre of the group and parameters $p_l$ and $p_b$ characterize
expansion (positive values) or compression (negative values) along
the $l$- or $b$-direction, respectively.  We solve the systems of
Eq.~\ref{pl} and Eq.~\ref{pb}  for all member stars of an
association with known {\it Gaia} DR2 proper motions to determine
the values of $p_l$ and $p_b$.

The observed specific velocities of expansion or compression,
$u_l$ and $u_b$, are computed as:

\begin{equation}
 u_l = p_l \; a, \quad\quad
 u_b =  p_b \; a.
\end{equation}

The motion of an OB-association as a whole with line-of-sight
velocity $V_r$ can also produce the effect of spurious
expansion/compression, which creates the expansion/compression
with the specific velocity $e_1$:

\begin{equation}
 e_1 = - V_r \; \frac{a}{r},
 \label{e1}
\end{equation}

\noindent where $a$ and $r$ are the specific radius  and
heliocentric distance of the OB-association, respectively. The
motion of a group toward the Sun ($V_r<0$) produces spurious
expansion ($e_1>0$) while the motion away from the Sun ($V_r>0$)
creates spurious compression ($e_1<0$). Generally, the observed
velocities, $u_l$ and $u_b$, must be corrected for this effect:

\begin{equation}
\tilde u_l = u_l - e_1, \quad\quad \tilde u_b =  u_b - e_1.
\end{equation}

\noindent Here we use the median values of line-of-sight
velocities of OB-associations, $V_r$   (Table~\ref{mass}), which
are derived from the velocities of individual stars taken from the
catalog by \citet{barbierbrossat2000}.

Table~\ref{par_expansion} lists the parameters of
expansion/compression, $p_l$ and $p_b$,  the velocities  $u_l$,
$u_b$, $e_1$, $\tilde u_l$, $\tilde u_b$ and their uncertainties
for 28 OB-associations containing more than 20 stars with {\it
Gaia} DR2 proper motions. We will discuss only velocities $\tilde
u_l$ or $\tilde u_b$ determined at a significance level greater
than $P>3\sigma$, which are underlined in
Table~\ref{par_expansion}.

\begin{figure*}
\resizebox{\hsize}{!}{\includegraphics{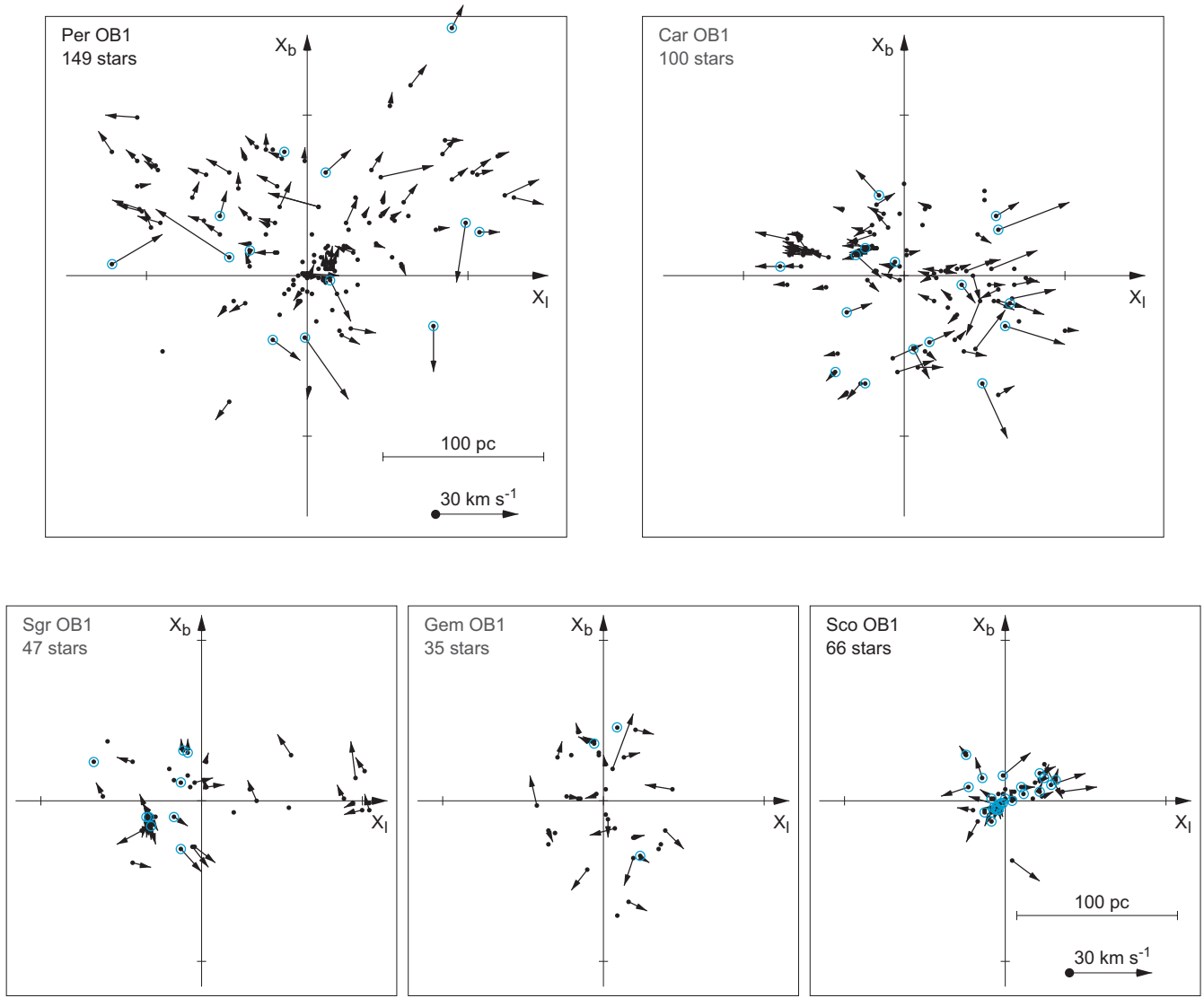}}
\caption{Distribution of observed relative velocities, $V_l'$ and
$V_b'$, in the Per OB1, Car OB1, Sgr OB1,  Gem OB1 and Sco OB1
associations.  All frames have the same scale. The velocities
$V_l'$ and $V_b'$ are determined with respect to the centre of the
association (Eqs~\ref{vl'} and ~\ref{vb'}). Stars with the
relative velocities $|V_l'|$ and $|V_b'|$ smaller than 3 km
s$^{-1}$ are shown as black circles without any vector. The axes
$X_l$ and $X_b$ are directed towards increasing values of Galactic
coordinates, $l$ and $b$, respectively. We can see conspicuous
expansion in these associations. Stars of spectral type O are
outlined by circles (colored blue in online article). O-type stars
can be seen to be distributed more or less uniformly among other
stars of OB-associations. } \label{ass4}
\end{figure*}

Figure~\ref{ass4} shows distribution of observed relative
velocities, $V_l'$ and $V_b'$, determined with respect to the
centre of the group:

\begin{equation}
 V_l' = v_l-v_{l0},
\label{vl'}
\end{equation}
\begin{equation}
 V_b' = v_b- v_{b0},
\label{vb'}
\end{equation}

\noindent in  five  OB-associations: Sgr OB1, Per OB1,  Gem OB1,
Car OB1 and Sco OB1, which demonstrate significant expansion in
both cases: in the observed relative velocities and after their
correction for the motion of the group as a whole. Four other
associations (Ori OB1, Cyg OB8, Cas OB2 and Mon OB2) will be
discussed below.

Here we excluded from consideration  17 member stars with the
absolute values of the relative velocities, $V_l'$ or $V_b'$,
greater than 50 km s$^{-1}$, because such large velocities cannot
be connected with expansion of OB-associations and are possibly
due to binary systems or runaway stars \citep[][see
Appendix]{fujii2011}.

Figure~\ref{ass4} also indicates the positions of O-type stars,
which are the youngest in OB-associations. We can see that  O-type
stars are distributed more or less uniformly  among other stars of
OB-associations.

Figure~\ref{ass4}  shows that the Sgr OB1 association includes
several groups of stars moving in opposite directions along the
$X_b$-axis. We do not observe here the expansion from one point as
it is, for example, in the Per OB1 or Car OB1 association. The
method of cluster analysis also shows that Sgr OB1 includes at
least two different groups \citep{melnik1995}.

As for Gem OB1,  it is just the exclusion of  five stars with the
error RUWE$>1.4$ from consideration  that allows us to find the
expansion inside this association.

\begin{figure*}
\resizebox{\hsize}{!}{\includegraphics{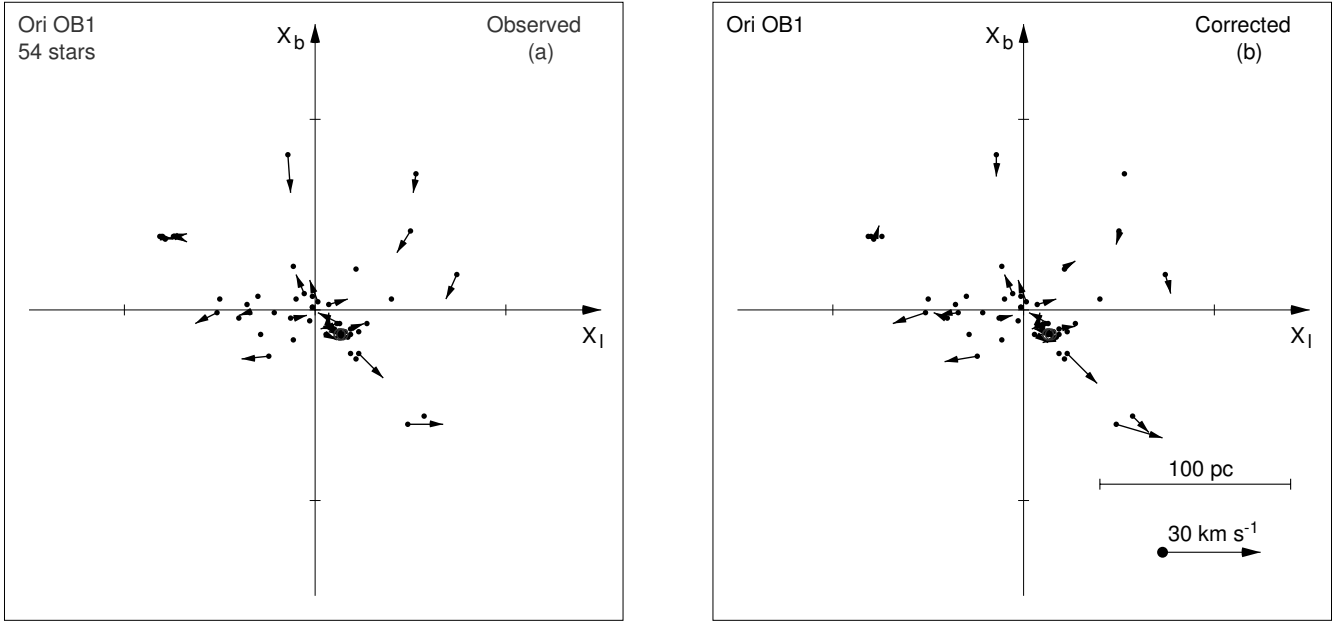}}
\caption{Distribution of observed, $V_l'$ and $V_b'$, and
corrected,  $V_{lc}'$ and $V_{bc}'$ (Eqs~\ref{vlc'} and
\ref{vbc'}), relative velocities in the Ori OB1 association. Stars
with the relative velocities $|V_l'|$ and $|V_b'|$ smaller than 3
km s$^{-1}$ are shown as black circles without any vector. The
axes $X_l$ and $X_b$ point in the $l$- and $b$-direction,
respectively. We can see small expansion along the $X_l$-axis in
the distribution of relative velocities corrected for the motion
of the association as a whole. } \label{ori}\end{figure*}

Figure~\ref{ori}  shows the distribution of  relative velocities
in the Ori OB1 association: Figure~\ref{ori}(a) displays the
observed relative velocities, whereas Figure~\ref{ori}(b) shows
the velocities corrected for the motion of the association as a
whole, $V_{lc}'$ and $V_{bc}'$:

\begin{equation}
V_{lc}' =  V_l' + V_r \; \frac{X_l}{r}, \label{vlc'}
\end{equation}
\begin{equation}
V_{bc}' =  V_b' + V_r \; \frac{X_b}{r}, \label{vbc'}
\end{equation}

\noindent where the pair ($X_l$, $X_b$) are the coordinates of a
star in the sky plane measured from the centre of OB-association
along $l$- and $b$-direction, respectively.  The velocity
corrections are zero at the centre of the group and increase
towards its periphery. The  velocity  of spurious expansion, $e_1$
(Eq.~\ref{e1}), in the Ori OB1 association is quite large in
absolute value, $e_1=-2.2$ km s$^{-1}$
(Table~\ref{par_expansion}), which is due to its location  near
the Sun ($r=0.4$ kpc) and its large line-of-sight velocity,
$V_r=+25.4$ km s$^{-1}$, resulting from a combined contribution
due to the solar motion to the apex (+13 km s$^{-1}$), the
rotation curve (+5 km s$^{-1}$) and the residual velocity (+7 km
s$^{-1}$). Note that the motion of the Ori OB1 association away
from the Sun produces spurious compression. Figure~\ref{ori} shows
that after the correction for the spurious compression, the
relative velocities demonstrate small expansion in the
$l$-direction.

\begin{figure}
\resizebox{\hsize}{!}{\includegraphics{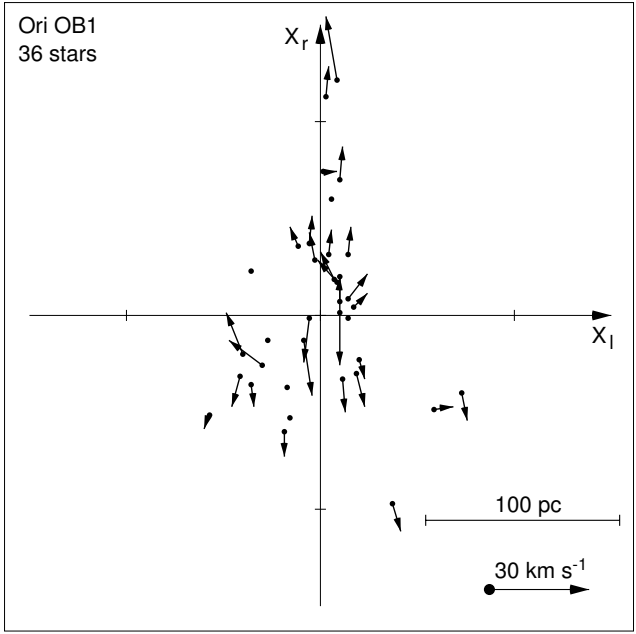}}
\caption{Distribution of relative velocities, $V_{lc}'$ and
$V_r'$, in the plane ($X_l$, $X_r$), where the axes $X_l$ and
$X_r$ point in the $l$-direction and  along the line of sight
connecting the Sun and the centre of the Ori OB1 association,
respectively. The Sun is at the bottom.  The velocity $V_r'$ is
the relative velocity along  the line of sight (Eq.~\ref{vr'}).
The velocity $V_{lc}'$ is the relative velocity in the
$l$-direction corrected for the motion of the association as a
whole (Eq.~\ref{vlc'}). Stars with the relative velocities
$|V_{lc}'|$ and $|V_r'|$ smaller than 3 km s$^{-1}$ are shown as
black circles without any vector. We can see conspicuous expansion
of the Ori OB1 association along the line of sight here.}
\label{ori_rad}\end{figure}

To check the expansion of the Ori OB1 association we analyze the
distribution of relative velocities in the  ($X_l$, $X_r$) plane,
where the $X_r$-axis  is directed along the line of sight
connecting the Sun and the centre of the Ori OB1 association
(Fig.~\ref{ori_rad}). The coordinate $X_r$ and the relative
velocity $V_r'$ of a star are derived from the {\it Gaia} DR2
parallax, $\pi$, and the  stellar line-of-sight velocity, $V_r$,
in the following way:

\begin{equation}
X_r =  \frac{1}{\pi} \, \cos \phi -  \frac{1}{\pi_0},
\end{equation}
\begin{equation}
V_r' =  V_r \, \cos \phi - V_{r0}, \label{vr'}
\end{equation}

\noindent where $\pi_0$ is the median parallax of stars of the Ori
OB1 association; $V_{r0}$ is the average stellar line-of-sight
velocity  projected onto the $X_r$-direction, and the angle $\phi$
is the heliocentric angle between the direction to the star and to
the centre of the association:

\begin{equation}
\phi = \arctan \frac{\sqrt{X_l^2+X_b^2}}{r},
\end{equation}

\noindent which  takes  values in the range $|\phi|<13^\circ$.

The parameter of expansion in the $X_r$-direction, $p_r$, is
determined by  solving the equations:

\begin{equation}
 V_r' =  p_r \, X_r.
\end{equation}

\noindent Figure~\ref{ori_rad} shows  conspicuous expansion of the
Ori OB1 association along the line of sight. The parameter of
expansion appears to be $p_r=105\pm29$ km s$^{-1}$ kpc  $^{-1}$
which corresponds to the specific velocity of expansion, $u_r$:

\begin{equation}
u_r =  p_r \, a,
\end{equation}

\noindent equal to $u_r=3.6\pm1.0$ km s$^{-1}$. Here we use the
line-of-sight velocities, $V_r$, and parallaxes, $\pi$,  measured
with the uncertainties less than 5 km s$^{-1}$ and 0.2 mas,
respectively, which are available  for 36 stars of the Ori OB1
association. The stellar line-of-sight velocities are taken from
the catalog by \citet{barbierbrossat2000}.

We found that the expansion/compression in the Cyg OB8, Cas OB2
and Mon OB2 associations relies  on the velocities of only a few
stars. The exclusion of $\sim10$ per cent of stars with {\it Gaia}
DR2 proper motions  from these associations (HD 191423 and HD
191778 from Cyg OB8, BD+63 1964 from Cas OB2, HD 262042, HD 47732
and HD 47777 from Mon OB2)  decreases the absolute value of
expanding/compressing velocity, $\tilde u_l$ or $\tilde u_b$, down
to a significance level of less than $2\sigma$.

\begin{figure*}
\resizebox{17 cm}{!}{\includegraphics{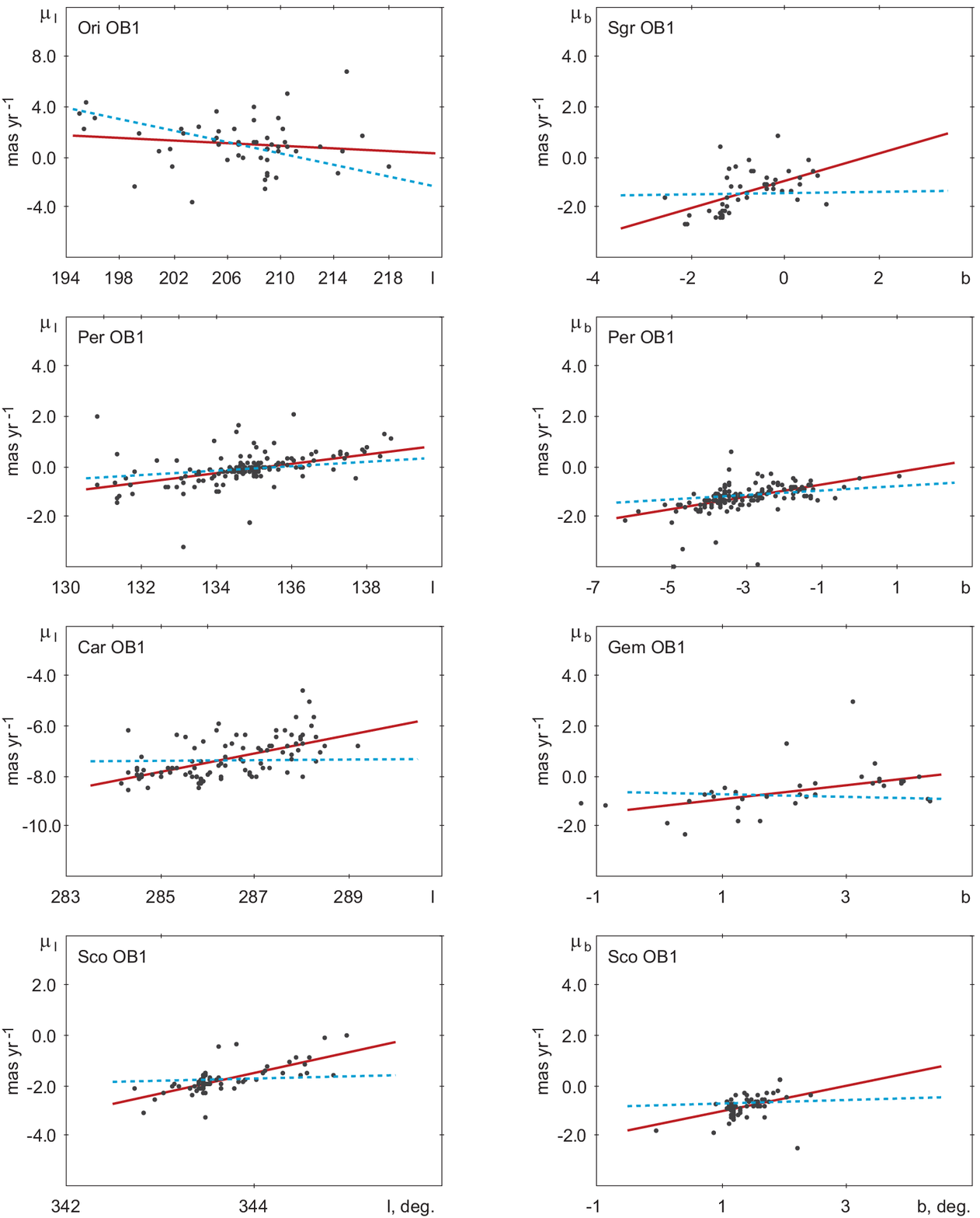}}
\caption{Dependence of  stellar proper-motion components, $\mu_l$
and $\mu_b$, on the corresponding coordinates, $l$ and $b$, in the
associations Sgr OB1, Per OB1, Ori OB1,  Gem OB1, Car OB1 and Sco
OB1. The solid lines (colored red in the online article) indicate
the linear fits of the dependencies  $\mu_l(l)$ and $\mu_b(b)$
determined from observational data while the dashed lines (colored
blue in the online article) show the correlation between $\mu_l$
and  $l$ or between $\mu_b$ and $b$, which appears due to the
motion of the association as a whole (Eq.~\ref{e1}). Positive
correlation between the proper motion and corresponding coordinate
means expansion while  negative correlation indicates compression.
In the Ori OB1 association, the spurious compression due to its
motion along the line of sight is considerably larger than that
derived from the observed data, which suggests the presence of a
physical expansion.} \label{lines}\end{figure*}

Figure~\ref{lines} shows the dependence of the  stellar proper
motions, $\mu_l$ and $\mu_b$, on the coordinates  $l$ and $b$,
respectively,  for the associations Sgr OB1, Gem OB1, Ori OB1, Per
OB1, Car OB1 and Sco OB1. Also shown is the  expected dependence
due to the motion of the association as a whole. We can see that
in all cases except Ori OB1, the correlation between $\mu_l$ and
$l$ or between $\mu_b$ and $b$ is positive and greater than the
correlation due to spurious expansion. Positive correlation
between the coordinate and corresponding proper motion suggests
expansion, whereas negative correlation means compression. The Ori
OB1 association is a special case -- here one expects to see
considerable negative correlation between $\mu_l$ and $l$ (the
dashed line) caused by the motion of the association away from the
Sun with $V_r=25$ km s$^{-1}$, but the observed dependence (the
red line) is much weaker than the expected one, which  indicates
the presence of some physical expansion  practically compensating
its spurious compression.

\begin{table*}
\caption{Expansion/compression  of OB-associations with {\it Gaia}
DR2 proper motions}
 \begin{tabular}{lccrrrrrrr}
 \\[-7pt] \hline\\[-7pt]
Name &    $p_l \quad \quad$ & $p_b \quad \quad$ & $a\quad$ &
$u_l\quad \; $ & $u_b\quad \;$ &  $e_1\quad \;\;$ &
$\tilde u_l\quad \; $ & $\tilde u_b\quad \;$ & $n_\mu$ \\
[2pt]
&  (km s$^{-1}$ kpc$^{-1}$) & (km s $^{-1}$ kpc$^{-1}$) & (kpc)$\;$  &  (km s$^{-1}$)&(km s$^{-1}$)&(km s$^{-1}$)&(km s$^{-1}$)&(km s$^{-1}$) &\\
  \\[-7pt] \hline\\[-7pt]
SGR OB5   & $  64\pm 63$ & $  -4\pm  25$ &0.068 &  $  4.4\pm 4.3$ & $ -0.3\pm 1.7$ & $ 0.43\pm 0.10$ & $  4.0\pm 4.3$ & $ -0.7\pm 1.7$ & 27 \\
SGR OB1   & $ -13\pm 11$ & $ 145\pm  36$ &0.037 &  $ -0.5\pm 0.4$ & $  5.3\pm 1.3$ & $ 0.29\pm 0.05$ & $ -0.8\pm 0.4$ & $  \underline{5.0\pm 1.3}$ & 47 \\
SER OB1   & $  -1\pm 33$ & $  44\pm  41$ &0.051 &  $ -0.1\pm 1.7$ & $  2.2\pm 2.1$ & $ 0.17\pm 0.12$ & $ -0.2\pm 1.7$ & $  2.1\pm 2.1$ & 33 \\
CYG OB3   & $ 249\pm109$ & $ 139\pm  67$ &0.024 &  $  6.1\pm 2.6$ & $  3.4\pm 1.6$ & $ 0.13\pm 0.02$ & $  5.9\pm 2.6$ & $  3.3\pm 1.6$ & 31 \\
CYG OB1   & $  -5\pm 68$ & $  25\pm  35$ &0.032 &  $ -0.2\pm 2.2$ & $  0.8\pm 1.1$ & $ 0.30\pm 0.02$ & $ -0.5\pm 2.2$ & $  0.5\pm 1.1$ & 62 \\
CYG OB9   & $ 162\pm 85$ & $ 113\pm 176$ &0.017 &  $  2.8\pm 1.5$ & $  2.0\pm 3.0$ & $ 0.35\pm 0.03$ & $  2.5\pm 1.5$ & $  1.6\pm 3.0$ & 22 \\
CYG OB8   & $ -46\pm 93$ & $ 178\pm  47$ &0.040 &  $ -1.8\pm 3.7$ & $  7.1\pm 1.9$ & $ 0.46\pm 0.05$ & $ -2.3\pm 3.7$ & $  \underline{6.7\pm 1.9}$ & 19 \\
CYG OB7   & $ -67\pm 69$ & $ 122\pm  51$ &0.051 &  $ -3.4\pm 3.5$ & $  6.2\pm 2.6$ & $ 0.76\pm 0.16$ & $ -4.1\pm 3.5$ & $  5.4\pm 2.6$ & 22 \\
CEP OB2   & $  72\pm 27$ & $  14\pm  22$ &0.046 &  $  3.4\pm 1.2$ & $  0.6\pm 1.0$ & $ 1.09\pm 0.06$ & $  2.3\pm 1.2$ & $ -0.4\pm 1.0$ & 45 \\
CAS OB2   & $ -32\pm 58$ & $ 121\pm  33$ &0.056 &  $ -1.8\pm 3.2$ & $  6.8\pm 1.8$ & $ 1.34\pm 0.05$ & $ -3.1\pm 3.2$ & $  \underline{5.4\pm 1.8}$ & 29 \\
CAS OB5   & $  54\pm 32$ & $  49\pm  22$ &0.043 &  $  2.3\pm 1.4$ & $  2.1\pm 1.0$ & $ 0.99\pm 0.02$ & $  1.3\pm 1.4$ & $  1.1\pm 1.0$ & 45 \\
CAS OB4   & $  95\pm 55$ & $  63\pm  20$ &0.070 &  $  6.7\pm 3.8$ & $  4.4\pm 1.4$ & $ 1.13\pm 0.05$ & $  5.6\pm 3.8$ & $  3.3\pm 1.4$ & 22 \\
CAS OB7   & $  74\pm 27$ & $  63\pm  35$ &0.041 &  $  3.0\pm 1.1$ & $  2.6\pm 1.4$ & $ 1.02\pm 0.00$ & $  2.0\pm 1.1$ & $  1.6\pm 1.4$ & 34 \\
CAS OB8   & $  51\pm 17$ & $  37\pm  19$ &0.037 &  $  1.9\pm 0.6$ & $  1.4\pm 0.7$ & $ 0.56\pm 0.02$ & $  1.3\pm 0.6$ & $  0.8\pm 0.7$ & 40 \\
PER OB1   & $  51\pm  9$ & $  65\pm  10$ &0.064 &  $  3.3\pm 0.6$ & $  4.2\pm 0.6$ & $ 1.52\pm 0.02$ & $  \underline{1.8\pm 0.6}$ & $  \underline{2.7\pm 0.6}$ &149 \\
CAS OB6   & $  89\pm 30$ & $ 185\pm  39$ &0.057 &  $  5.1\pm 1.7$ & $ 10.6\pm 2.2$ & $ 1.39\pm 0.05$ & $  3.7\pm 1.7$ & $  \underline{9.2\pm 2.2}$ & 28 \\
CAM OB1   & $  36\pm 14$ & $ -21\pm  17$ &0.079 &  $  2.8\pm 1.1$ & $ -1.6\pm 1.3$ & $ 1.08\pm 0.14$ & $  1.7\pm 1.1$ & $ -2.7\pm 1.3$ & 41 \\
AUR OB1   & $  11\pm 14$ & $  -8\pm  15$ &0.072 &  $  0.8\pm 1.0$ & $ -0.6\pm 1.1$ & $ 0.13\pm 0.17$ & $  0.7\pm 1.0$ & $ -0.7\pm 1.1$ & 31 \\
ORI OB1   & $ -15\pm 14$ & $ -57\pm  16$ &0.035 &  $ -0.5\pm 0.5$ & $ -2.0\pm 0.5$ & $-2.21\pm 0.09$ & $  \underline{1.7\pm 0.5}$ & $  0.2\pm 0.5$ & 54 \\
GEM OB1   & $  10\pm 34$ & $  77\pm  26$ &0.042 &  $  0.4\pm 1.4$ & $  3.2\pm 1.1$ & $-0.55\pm 0.03$ & $  1.0\pm 1.4$ & $  \underline{3.8\pm 1.1}$ & 35 \\
MON OB2   & $-215\pm 45$ & $  36\pm  66$ &0.037 &  $ -7.9\pm 1.6$ & $  1.3\pm 2.4$ & $-0.70\pm 0.07$ & $ \underline{-7.2\pm 1.6}$ & $  2.0\pm 2.4$ & 23 \\
VELA OB1  & $  26\pm 22$ & $ -29\pm  16$ &0.057 &  $  1.5\pm 1.2$ & $ -1.7\pm 0.9$ & $-0.89\pm 0.03$ & $  2.4\pm 1.2$ & $ -0.8\pm 0.9$ & 42 \\
CAR OB1   & $ 101\pm 14$ & $  37\pm  14$ &0.064 &  $  6.5\pm 0.9$ & $  2.4\pm 0.9$ & $ 0.16\pm 0.03$ & $  \underline{6.3\pm 0.9}$ & $  2.2\pm 0.9$ &100 \\
CAR OB2   & $  59\pm 59$ & $  83\pm  27$ &0.028 &  $  1.6\pm 1.6$ & $  2.3\pm 0.8$ & $ 0.13\pm 0.02$ & $  1.5\pm 1.6$ & $  2.2\pm 0.8$ & 48 \\
CRU OB1   & $  35\pm 31$ & $  33\pm  16$ &0.040 &  $  1.4\pm 1.2$ & $  1.3\pm 0.6$ & $ 0.11\pm 0.02$ & $  1.3\pm 1.2$ & $  1.2\pm 0.6$ & 64 \\
CEN OB1   & $ -18\pm 18$ & $  13\pm  11$ &0.068 &  $ -1.2\pm 1.2$ & $  0.9\pm 0.7$ & $ 0.67\pm 0.06$ & $ -1.9\pm 1.2$ & $  0.2\pm 0.7$ & 84 \\
ARA OB1A  & $  31\pm 31$ & $  82\pm  39$ &0.046 &  $  1.4\pm 1.4$ & $  3.7\pm 1.8$ & $ 1.51\pm 0.13$ & $ -0.1\pm 1.4$ & $  2.2\pm 1.8$ & 42 \\
SCO OB1   & $ 224\pm 29$ & $ 142\pm  38$ &0.013 &  $  3.0\pm 0.4$ & $  1.9\pm 0.5$ & $ 0.25\pm 0.02$ & $  \underline{2.8\pm 0.4}$ & $\underline{1.7\pm 0.5}$ & 66 \\

\hline \end{tabular} \label{par_expansion}
\end{table*}


\subsection{Kinematic ages of OB-associations}

We can make crude estimates of stellar ages on the basis of their
spectral types (effective temperatures) and luminosities using
available stellar models and various photometric calibrations. The
ages of O8 stars of all luminosity classes must be smaller than 5
Myr, while B0 stars must be younger than  10 Myr
\citep{bressan2012}. So the ages of  OB-associations are supposed
to be 5--10 Myr.  However, the low-mass stars born in the same
molecular cloud can be conspicuously older, some estimates of
their ages are as old as 20--50 Myr \citep{pecaut2016,
cantat2019}. In this context it is interesting to compare the ages
of OB-associations derived from stellar models (the so-called
stellar ages) with their kinematic ages, i.e. the time instants in
the past when the group had minimal size.  There are two classical
methods of the determination of kinematic ages. One method derives
the kinematic ages from $p_l$ and $p_b$, while another is based of
the backtracing the positions of individual stars.
\citet{brown1997} tested both methods and found that the first
method overestimates the kinematic age while the second method, on
the contrary, underestimates it.

Table~\ref{age} presents the estimates of the kinematic ages
determined by two different ways. The values $T_l$ and $T_b$ are
computed by the first method through the values of $p_l$, $p_b$,
$e_1$ and $a$:

\begin{equation}
 T_l=[(p_l-\frac{e_1}{a})\, f_v]^{-1},
\label{T_l}
\end{equation}
\begin{equation}
 T_b=[(p_b-\frac{e_1}{a})\, f_v]^{-1},
\label{T_b}
\end{equation}

\noindent where factor $f_v=1.023 \cdot 10^{-3}$ transforms
velocities in units of km s$^{-1}$ into kpc Myr$^{-1}$. The
parameters of expansion, $p_l$ and $p_b$, were corrected for the
motion of the association as a whole.

The second method is based on the positions and proper motions of
individual stars.  We determine the positions of member stars in
the past using their present-day coordinates and velocities. Here
we suppose that stars in an association start their expansion at
one moment in the past but with different velocities. The observed
velocities computed with respect to the centre of the association,
$v_l'$ and $v_b'$, were corrected for the motion of the
association as a whole:

\begin{equation}
 x(t) = x_0 - (v_l'+V_r \; \frac{x_0}{r})\, t,
\label{past_x}
\end{equation}
\begin{equation}
 y(t) = y_0 - (v_b'+V_r \; \frac{y_0}{r})\, t,
\label{past_y}
\end{equation}

\noindent The selection of the most compact part of association
allows us to mitigate changes due to inclusion or exclusion of
individual stars. Only cases with well-defined expansion are
considered.

\begin{figure*}
\resizebox{\hsize}{!}{\includegraphics{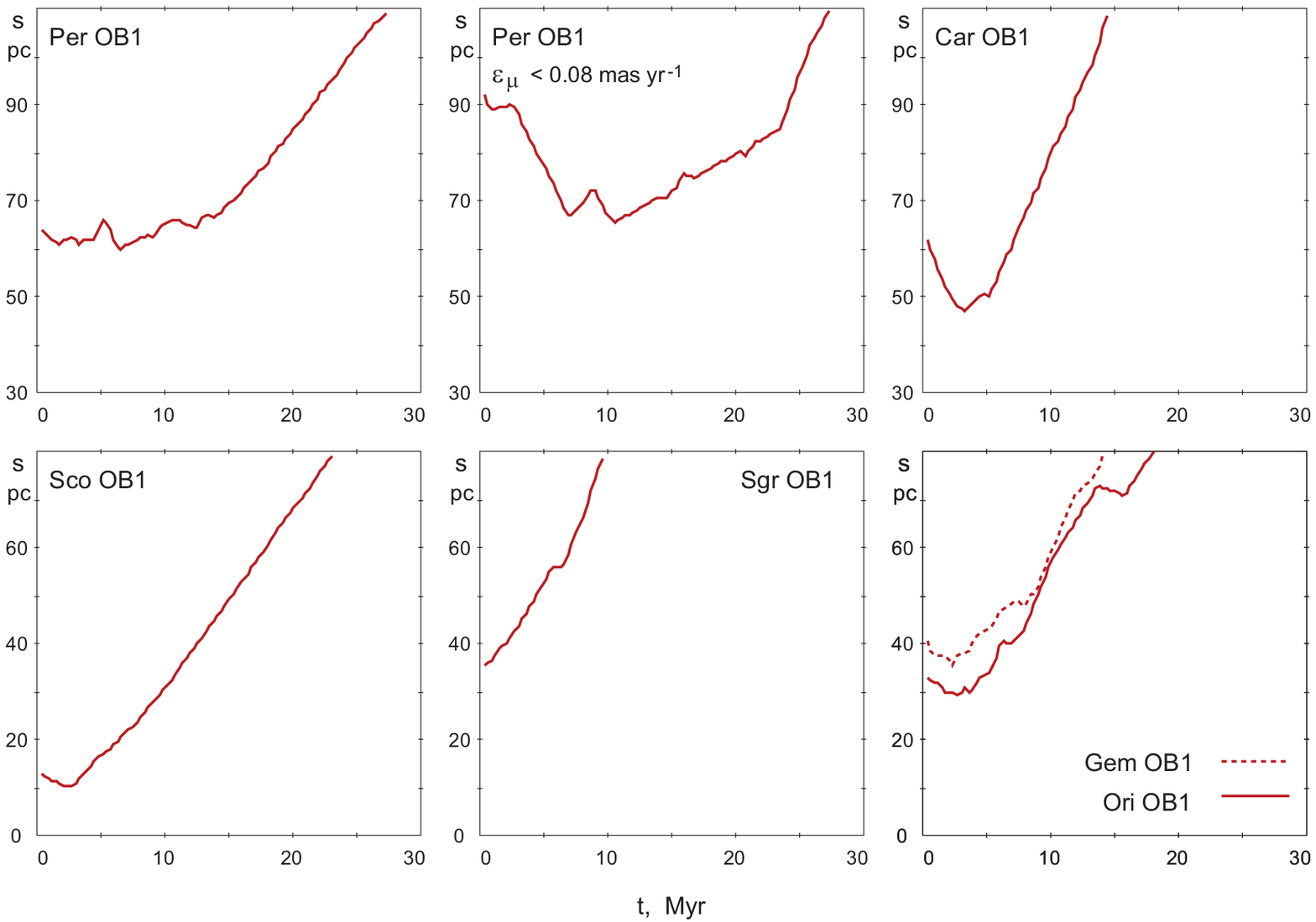}}
\caption{Dependence of the size $s$ of OB-association on the time
$t$ in the past.  Here $s$  is the radius of association
containing central 68 per cent of its star-members with known Gaia
DR2 proper motions. In the case of the Per OB1 association we also
show separate plot for the sample of stars with the most precise
proper motions ($\varepsilon_{\mu l}<0.008$ mas yr$^{-1}$ and
$\varepsilon_{\mu b}<0.008$ mas yr$^{-1}$). Only cases with
well-defined parameters of expansion are considered.}
\label{curves}\end{figure*}

Figure~\ref{curves} shows the dependence of the size of
OB-association, $s$, on the time $t$ in the past. The value $s$ is
the radius of the association containing central 68 per cent of
its members with known {\it Gaia} proper motions.  Note that the
second method is very sensitive to the errors in proper motions of
individual stars. So  we select stars with the most precise proper
motions in the Per OB1 association and calculate  function $s(t)$
for this sample as well. The new sample includes  stars with
$\varepsilon_{\mu l}<0.008$ mas yr$^{-1}$ and $\varepsilon_{\mu
b}<0.008$ mas yr$^{-1}$ leaving 38 stars  in the Per OB1
association.

Figure~\ref{curves} shows  that function $s(t)$  has a minimum in
some cases  and a plateau with subsequent growth in others. In the
Per OB1 association the sample of stars with the most precise
proper motions shows minimum on the curve $s(t)$ while the sample
including all stars with {\it Gaia} DR2 proper motions gives a
plateau. The selection of stars with the most accurate proper
motions does not change the results in the Sgr OB1, Car OB1 and
Sco OB1 associations, so we don't demonstrate them separately.
Note that the average uncertainty in determination of {\it Gaia}
DR2 proper motions in the Ori OB1 association is $\varepsilon_{\mu
l}=0.230$ mas yr$^{-1}$, and the criteria considered  leaves no
stars here.

Kinematic ages of OB-associations obtained by the second method,
$T^*$, correspond to the minimum on the curves $s(t)$. Note that a
considerable difference between the minimum  and present-day sizes
of an association:

\begin{equation}
\xi=s(T^*_l)/s(0), \label{size}
\end{equation}

\noindent is observed only in the Per OB1 and Car OB1 associations
equal to   $\xi=0.69$ and $\xi=0.73$, respectively. The minimal
radii of OB-associations corresponding to the plateau or to a
minimum of functions $s(t)$ are: 67 pc (Per OB1), 47 pc (Car OB1),
10 pc (Sco OB1), 35 pc (Sgr OB1), 29 pc (Ori OB1). These values
are located in the interval 10--100 pc corresponding to the
expected sizes of giant molecular clouds \citep{sanders1985}.

Table~\ref{age} lists the kinematic ages $T^*$  and their errors
calculated from the  uncertainties in proper motions which cause
the uncertainties in the determination of sizes of
OB-associations, $\Delta s$, proportional to the time interval in
the past:

\begin{equation}
\Delta s =  4.74\, r \, \varepsilon_{\mu} \; T^* f_v,
\end{equation}

\noindent where $\varepsilon_{\mu}$ is the average uncertainty in
determination of proper motions  in $l$- and $b$-directions. The
values of  $\Delta s$ are followings: 4 pc (Per OB1, sample of
stars with the most precise proper motions), 2 pc (Car OB1, Sco
OB1, Gem OB1), 1 pc (Sgr OB1, Ori OB1).

In the case of Sgr OB1  we can give only upper limit for the value
of $T^*$. The values of $T^*$ do not exceed 4 Myr except the case
of the Per OB1 association where it amounts to
$T^*=10.6^{+2.6}_{-3.9}$ Myr.

Table~\ref{age} suggests that the kinematic ages, $T_l$ and $T_b$,
obtained by the first method are always greater than corresponding
estimates derived  using the second method, $T^*$,  which agrees
with the results by \citet{brown1997}. The first method
(Eqs.~\ref{T_l} and \ref{T_b}) is less sensitive to the errors in
proper motions of individual stars: several erroneous proper
motions cannot spoil the picture of overall expansion
(Fig.~\ref{lines}), but this method assumes that the
OB-association expands from a point  or from a very small region
which can be incorrect  in some cases. The second method
(Eqs.~\ref{past_x} and \ref{past_y}) is more direct: it does not
involve  additional assumptions about the initial size of the
group, but it is more sensitive to the errors in proper motions of
individual stars which can completely wash out the expansion
(Fig.~\ref{curves}).

Our estimates of the kinematic ages of OB-associations, $T^*$,
obtained by the second method are  smaller than 5 Myr in the Sgr
OB1, Gem OB1, Ori OB1, Car OB1 and Sco OB1  associations which
agrees with the following star-formation scenario: first OB-stars
form inside a giant molecular cloud and only then a group of these
stars starts its expansion. In the case of the Per OB1, the value
of $T^*=10.6^{+2.6}_{-3.9}$ is only marginally consistent with
this sequence of events. Possibly, the Per OB1 association started
its expansion with a larger  velocity than its present-day value,
resulting in a smaller  kinematic age (see also section 3.5).

\begin{table}
\caption{Kinematic ages of OB-associations}
 \begin{tabular}{lcc|c}
 \\[-7pt] \hline\\[-7pt]
Name &$T_l$&$T_b$&$T^*$\\
& Myr & Myr & Myr \\
  \\[-7pt] \hline\\[-7pt]
Sgr OB1   &   --        & $7.1^{+0.6}_{-1.5}$ & $<1$    \\
[+5pt]
Per OB1   &      $36^{+17}_{-8}$ & $24^{+7}_{-5}$  &           $10.6^{+2.6}_{-3.9}$          \\
[+5pt]
Ori OB1   &  $20.4^{+7.2}_{-4.6}$ &  --  &  $2.7\pm0.3$  \\
[+5pt]
Gem OB1   &  $10.9^{+1.4}_{-2.4}$ &  --  &  $2.5^{+0.1}_{-0.6}$  \\
[+5pt]
Car OB1   &  $9.9^{+1.6}_{-1.2}$ &  --  &  $3.4^{+0.6}_{-0.8}$   \\
[+5pt]
Sco OB1   &  $4.8^{+0.8}_{-0.6}$ &  $8.0^{+2.5}_{-1.9}$  & $2.1\pm1.4$    \\
[+5pt] \hline \end{tabular} \label{age}
\end{table}

\subsection{Shell-like structure in the distribution of stars  in the Per OB1
association}

\begin{figure*}
\resizebox{\hsize}{!}{\includegraphics{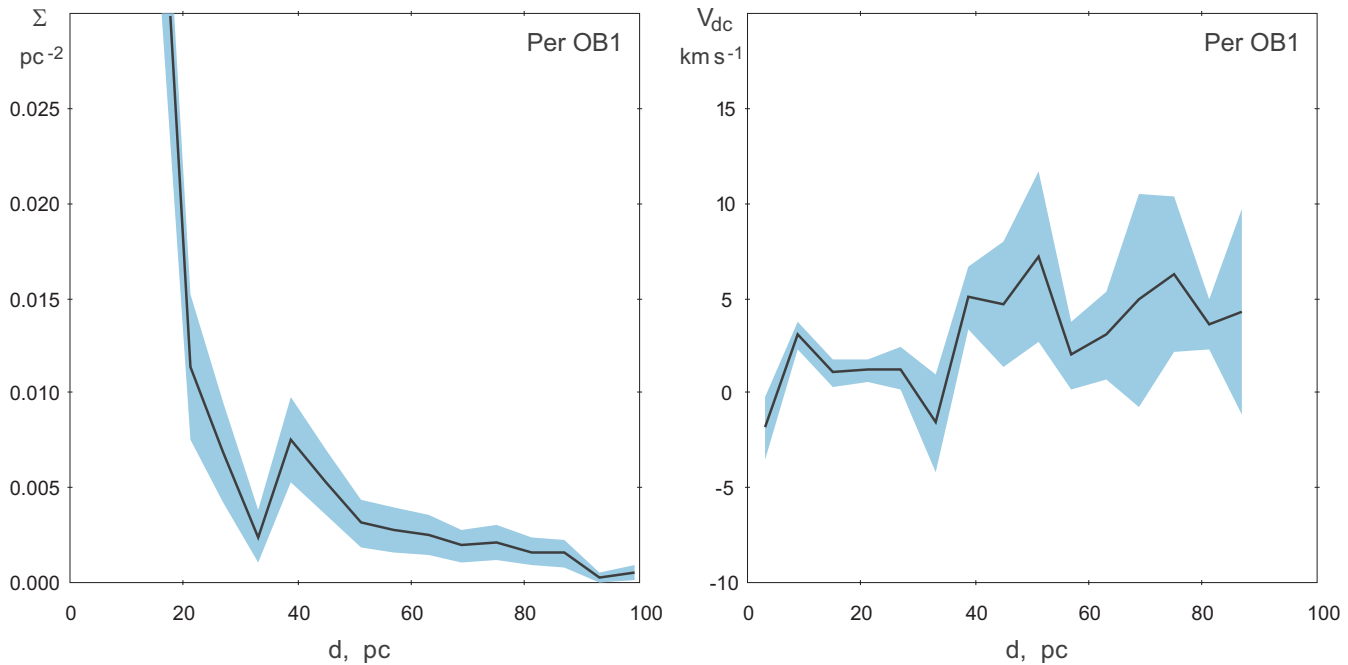}} \caption{(a)
Distribution of OB-stars inside the Per OB1 association averaged
in thin annuli.  The distribution of stars in the sky-plane was
divided into  6 pc-wide  annuli of radius $d$ with the centres
coincident with the centre of the association. The horizontal and
vertical axes measure  the distance from the centre, $d$, and the
surface density of stars, $\Sigma$, respectively.  The
uncertainties due to Poisson noise are shown in gray (colored blue
in the online article). We can see the existence of a secondary
maximum  at the distance $40$ pc from the centre of the
association. (b) Dependence of the velocity of expansion,
$V_{dc}$, averaged in thin annuli on the distance $d$. The
velocity $V_{dc}$ is the relative corrected  velocity in the
direction connecting the centre of the association with the star.
The velocity of expansion at the distance of the secondary maximum
appears  to be $5.0\pm1.7$ km s$^{-1}$. The scatter due to
root-mean-square errors in determination of average velocities in
each annulus is shown in gray (colored blue in the online
article).} \label{shell}\end{figure*}

The distribution of stars in two large expanding OB-associations,
Per OB1 and Car OB1, suggests the presence of a shell-like
structure with a  density maximum located between the central part
and periphery (Fig.~\ref{ass4}). To check this hypothesis we
consider the variations of the surface density of OB-stars,
$\Sigma$, with the distance, $d$,  from the centre of the
association (Fig.~\ref{shell}a). We subdivided the distribution of
stars in the sky-plane inside the association   into 6 pc-wide
annuli of radius $d$ with the centres coincident with the centre
of the association. The average surface density of stars at the
distance $d$ from the centre is the ratio $\Sigma=N/S$, where $N$
is the number of stars in the annulus and $S$ is its area.

Figure~\ref{shell}(a) shows the presence of a secondary maximum on
the curve $\Sigma(d)$ for the  Per OB1 association located at the
distance of 40 pc from its centre.   The secondary maximum is due
to the presence of a minimum on the density curve between the
central concentration and the periphery. The statistical
significance of the secondary maximum  is determined by the errors
due to Poisson noise in the maximum and minimum density
distribution, $\sigma_1$ and $\sigma_2$, respectively:

\begin{equation}
S=\frac{\Sigma_{max} - \Sigma_{min}}{\sigma_1 +\sigma_2},
\end{equation}

\noindent which gives the significance  level of $P \sim
1.4\sigma$. We find a similar shell-like structure in the Car OB1
association but it has  lower statistical significance and we do
not consider it here.

To understand the nature of the secondary maximum we build the
distribution of velocities inside the association.
Figure~\ref{shell}(b) shows the variation of the average expansion
velocity, $V_{dc}$, of OB-stars with the distance, $d$, from the
centre of the association. The velocity $V_{dc}$ of a star is the
component of its relative corrected velocity, ($V_{lc}'$,
$V_{bc}'$) (Eqs.~\ref{vlc'} and  ~\ref{vbc'}), directed along the
radius-vector, $\overrightarrow{d}$, connecting the centre of the
association with the star in the sky plane. We can see that  the
velocity $V_{dc}$ at the distance of the shell, $d=40$ pc, amounts
to the value of $V_{dc}=5.0\pm1.7$ km s$^{-1}$.  Thus, we can
speak about the expanding shell of stars here.

There are two different scenarios of the formation of expanding
OB-associations. The first  one considers  the  expansion of young
stellar group  caused by  the gas loss in its parent molecular
cloud \citep{hills1980,kroupa2001,boily2003a, boily2003b,
vine2003}. The second scenario includes two episodes  of star
formation: the first generation of massive stars creates  an
expanding gas shell while the second generation of stars, which we
observe now,  forms from the molecular gas collected by the
expanding shell. The radii of the shells are supposed to be
30--150 pc and the velocity of expansion $V_{sh}$ must be less
than $V_{sh}<15$ km s$^{-1}$ \citep[][and
others]{castor1975,weaver1977,elmegreen1977,lozinskaya1988,
lozinskaya1999}. There is a difference between the formation of
OB-stars in an expanding shell, when stars acquire the expansion
velocity at the time of their birth, and the expansion of
OB-association due to the gas loss in the parent molecular cloud,
when stars are born in the turbulent but unexpanding gas medium
and begin their motion outwards  due to a sudden lack of
gravitational force in the centre of the group.

The formation of the Per OB1 association from the molecular gas
accumulated by the shell seems to be unlikely. There are two
arguments against this scenario.  First, we do not observe any
difference in ages of OB-stars located in the central part  and in
the expanding shell in the Per OB1 association (Fig.~\ref{ass4}).
Second, the symmetry in the distribution of stars with respect to
the centre of the association. If stars were born in an expanding
shell then they would  have been formed in gravitationally
isolated segments of the shell which must have a bit different
density, so the formation of massive stars throughout the shell in
a small time interval looks questionable.

We suppose that the expansion of the Per OB1 association is caused
by the gas loss in its parent molecular cloud  due to thermal
pressure of HII regions. The average radius and expansion velocity
of the shell are $d=40$ pc  and $V_{dc}=5.0\pm1.7$ km s$^{-1}$,
respectively. It means that  the Per OB1 association started its
expansion $T=8^{+4}_{-2}$ Myr ago.  However, the existence of
massive O-type stars in the expanding shell, whose ages do not
exceed 5 Myr (Fig.~\ref{ass4}), suggests that the expansion of the
Per OB1 association must have started with a larger expanding
velocity and its kinematical age must be smaller than 5 Myr.

Note that the age of the double stellar cluster $h$ and $\chi$
Persei located in the centre of the Per OB1 association is
supposed to be 10--13 Myr \citep{slesnick2002,dias2002}, which
suggests that the formation of their stars could have started 5--8
Myr before the formation of present-day O-stars in the Per OB1
association. Probably,   stars of the clusters $h$ and $\chi$
Persei formed during the time interval when the parent molecular
cloud was compressing due to its own gravity but the mean gas
density had not  reached its maximal value yet. Simulations show
that during the global collapse of the cloud    the average Jeans
mass is decreasing and the fraction of gas mass involved in the
instantaneous star formation is increasing, so the initial mass
function  is expected to be sufficiently sampled to produce
massive stars which then begin to erode the cloud
\citep{colin2013,zamora2012}. Probably, the giant molecular cloud
in which  the Per OB1 association formed had very dense central
part which allowed it to produce  a lot of massive stars during a
small time interval.

\section{Discussion and conclusions}

We studied the velocities in the sky plane, $v_l$ and $v_b$,
inside 28 OB-associations including more than 20 {\it Gaia} DR2
stars. The average  velocity dispersion inside 28 associations is
$\overline{\sigma_v}=4.5\pm0.3$ km s$^{-1}$, which exceeds the
corresponding value derived from {\it Gaia} DR1 proper motions,
$\overline{\sigma_v}=3.5\pm0.5$ km s$^{-1}$, at the significance
level of $P\sim 1\sigma$.  The greater value of $\sigma_v$
obtained with {\it Gaia} DR2 proper motions can be both due to
statistical effects  and due to shorter time base-line, which
makes the results more sensitive to all sorts of systematic
effects.

\citet{kounkel2018} found that the line-of-sight velocities of
stars inside the Ori OB1 association  obtained with the
uncertainty less than 1 km s$^{-1}$  and corrected for the motion
of the Sun toward the standard apex  lie in the range 0--15 km
s$^{-1}$ which is consistent with the velocity dispersion 3--5 km
s$^{-1}$.

We used the velocity dispersion $\sigma_v$ inside OB-associations
and their radii $a$ to  calculate the  virial masses of
OB-associations, which are equal to the masses of their parent
giant molecular clouds. The median virial mass computed for 28
OB-associations  is $M_{vir}=8.9 \times 10^5$ M$_\odot$. We also
determined the stellar masses of OB-associations. Here we used the
power-law distribution of  stars over  masses by
\citet{kroupa2002} calibrated through  the number of massive stars
in OB-association. The median  stellar mass of OB-associations is
$M_{st}=8.1 \times 10^3$ M$_\odot$. The  stellar to virial mass
ratio determines the average  star-formation efficiency inside the
giant molecular cloud, which has the  median value  of
$\epsilon=1.2$ per cent.

We found that  the  Sgr OB1, Per OB1,  Ori OB1,  Gem OB1, Car OB1
and Sco OB1 associations are expanding   at a significance level
of $P>3\sigma$. The expansion of the Per OB1, Car OB1 and Sgr OB1
associations first  have been found  with {\it Gaia} DR1 proper
motions \citep{melnik2017}. {\it Gaia} DR2 data confirm their
expansion and suggest   three new expanding associations: Ori OB1,
Gem OB1 and Sco OB1. The Ori OB1 association shows expansion only
after the correction of velocities $v_l$ and $v_b$ for the
line-of-sight motion of the association as a whole, it also
displays the expansion along the line of sight.  The expansion of
the Gem OB1 association was found only after the exclusion of five
stars with the error RUWE$>1.4$ from consideration.  Only two
associations (Per OB1 and Sco OB1) show well-defined expansion in
both the $l$- and $b$-direction, the other associations
demonstrate expansion at a significance level $P>3\sigma$ in one
direction only.

We gave evidence in favour of  the expanding stellar shell in the
Per OB1 association with the radius of $d=40$ pc and the expansion
velocity  of $V_{dc}=5.0\pm1.7$ km s$^{-1}$, which suggests that
the expansion of OB-stars started $T=8^{+4}_{-2}$   Myr ago.
However, the existence of massive O-type stars in the expanding
shell with the ages less than 5 Myr suggests that the expansion of
the Per OB1 association started with a larger expanding velocity,
resulting in a smaller kinematic age.

We can see that the expansion of OB-associations is quite a rare
event: it is observed only  in  6 from 28 associations considered,
i.e. the frequency of this event  is $\sim 21$ per cent. There are
two possible reasons of such a low frequency of expanding
OB-associations. First, it is the lack of precise proper motions
for stars of OB-associations so the expansion is merely washed out
by noise. Second, it is a physical reason -- only the most dense
giant molecular clouds  can produce stellar groups expanding from
one centre, while less dense  clouds produce several groups inside
one cloud  each of which is expanding from its  centre, so the
total velocity distribution looks chaotic.

We found  a strong correlation between the re-normalised unit
weight errors (RUWE) and the  relative velocities (Eqs~\ref{vl'}
and \ref{vb'}) of member stars determined with respect to the
centre of OB-association. Note that among 33 stars moving with the
relative velocities more than 50 km s$^{-1}$ the fraction of stars
with RUWE$>1.4$ amounts to 48 per cent but the expected fraction
is only 9 per cent.

\section{acknowledgements}

We thank the anonymous referee  for fruitful discussion. This work
has made use of data from the European Space Agency (ESA) mission
{\it Gaia} (https://www.cosmos.esa.int/gaia), processed by the
{\it Gaia} Data Processing and Analysis Consortium (DPAC,
https://www.cosmos.esa.int/web/gaia/dpac/consortium). Funding for
the DPAC has been provided by national institutions, in particular
the institutions participating in the {\it Gaia} Multilateral
Agreement. This research has made use of the VizieR catalogue
access tool, CDS, Strasbourg, France. The original description of
the VizieR service was published  by \citet{ochsenbein2000}.


\subsection{Appendix}

Here is the list of  17 stars of OB-associations  with the
absolute values of the relative velocities, $V_l'$ or $V_b'$,
greater  than 50 km s$^{-1}$, which were excluded from our
consideration in the study of the expansion of OB-associations
(section 3.3): HD 163065 (Sgr OB5),  HD 166937A (Sgr OB1),  HD
172488 (Sct OB2), HD 192445 (Cyg OB3), HD 192281 (Cyg OB8),  HD
212043 (Cep OB2),  HDE 240331 (Cas OB2),  BD +59 70 (Cas OB4), BD
+62 68 (Cas OB4), HILTNER 74 (Cas OB7), BD +60 261 (Cas OB8),  HD
12323 (Per OB1), BD +60 512 (Cas OB6),  HD 73634 (Vela OB1), HD
94024 (Car OB1),  HD 102248 (Cru OB1), HD 112272 (Cen OB1).

\end{document}